\documentclass[a4paper,11pt]{article}
\pdfoutput=1 

\usepackage{jinstpub} 

\title{\boldmath Long term operation with the DarkSide-50 detector}

\author[1]{N. Canci\note{Corresponding author.}}

\affiliation{Department of Physics, University of Houston\\Houston, TX 77204, U.S.A}
\affiliation{INFN - Laboratori Nazionali del Gran Sasso\\Assergi AQ 67010, Italy}

\emailAdd{nicola.canci@lngs.infn.it}

\abstract{DarkSide is a staged experimental project based on radiopure argon aiming at direct dark matter detection. The DarkSide-50 (DS-50) detector is currently operating underground at the Gran Sasso National Laboratory.
DS-50 detector is a dual-phase, 50~kg, liquid argon time-projection-chamber readout by 38 cryogenic PMTs (Hamamatsu R11065), surrounded by an active liquid scintillator veto and contained in a water Cherenkov detector acting as a muon veto.\\
DS-50 has been been operating continuously since 2013, first with atmospheric argon and subsequently filled in 2015 with argon from an underground source, allowing a reduction of the $^{39}$Ar isotope by more than a factor 1000.\\
Features of the DS-50 detector are described, long term operations and stability are reported and its performances in scintillation light detection discussed.
Results on dark matter searches obtained with DarkSide-50 detector are briefly reported.}

\keywords{Liquid argon; Noble liquid detectors (scintillation, ionization, double-phase); Dark Matter detectors (WIMPs, axions, etc.)}

\collaboration[c]{on behalf of DarkSide-50 Collaboration}

\proceeding{LIDINE2019\\
  August 28-30, 2019\\
  University of Manchester}

\begin{document}
\maketitle
\flushbottom


\section{Introduction}
\label{sec:intro}

The existence of an unknown nature form of matter, called {\it{dark matter}}, has been shown by different astronomical and cosmological evidences, indicating that only a little fraction of the matter of the Universe is composed of ordinary baryonic matter.
Proofs of the possible presence of dark matter have been found coming from several and independent investigations in different fields of research. 
A favored hypothesis that explains these observations is that dark matter is made of Weakly Interacting Massive Particles (WIMPs).
A great effort is then currently devoted to find some evidence of these new particles \cite{bertone1, bertone2, schumann, young}.\\
Direct dark matter search experiments look for an excess of signals in an underground, low background environment; different detectors have been built for this purpose, with the aim to measure small energy depositions (1$\div$100~keV).
The use of liquified noble gases (neon, argon and xenon) as sensitive medium is one of the most promising direct detection techniques to perform this search. 
Argon in particular, due to its features as high scintillation photon yield, powerful background rejection through efficient event discrimination methods, ease of purification and high abundance at reasonable cost, represents an ideal medium for the WIMPs detection.\\ 
In direct dark matter search scenario a leading role is played by the DarkSide (DS) project based on ultrapure depleted argon.


\section{The DarkSide program}
\label{sec:ds_program}

The DarkSide program has been designed with the aim of direct dark matter detection via WIMP-nucleus scattering in liquid argon (LAr) at Gran Sasso National Laboratory, using a dual-phase LAr Time Projection Chamber (TPC), with scalable, zero-background technology \cite{warp, ds_prog}.\\
The two-phase (liquid-gas) technology is based on the simultaneous detection of both signals produced by ionization events in liquid argon: free electron charge and scintillation light.
In fact, particles interacting in LAr create free electrons and excited Ar molecular states which produce scintillation radiation through de-excitation processes. The two processes are complementary and their relative weight depends on the strength of the electric field applied to the active LAr volume \cite{doke, kubota1, kubota2}.
The main feature of the LAr technology in the direct dark matter search is the capability to efficiently separate nuclear recoil (NR) events, as expected from WIMP interactions, from electron recoil (ER) events.\\
In general, the largest challenge in searching for dark matter is the suppression of the rate of background events to below the very low WIMP interaction rates (a few events per ton$\times$year) to which current dark matter experiments are sensitive.
In the case of the LAr technology, one of the main drawback is the presence of the cosmogenically produced $\beta$-emitter (with a half-life of 269 years) radioactive isotope of $^{39}$Ar in natural atmospheric argon (AAr) with an intrinsic activity of $\sim$1~Bq/kg, end-point of the $\beta$ spectrum at 565~keV and mean energy at 220~keV \cite{benetti}. 
The low energy electrons show an energy in the range of the interest for a WIMP search and can become a problem for larger detectors.
Therefore the reduction of $^{39}$Ar content represents a crucial requirement for large scale LAr dark matter detector. To this purpose a particular effort has been dedicated in procuring argon from underground reservoirs: underground argon (UAr), protected from cosmic ray activation, is significantly reduced in $^{39}$Ar content at least of 2$\div$3 orders of magnitude respect to the AAr \cite{acosta, calaprice}.\\
The DarkSide project foresees a multi-stage approach and after a first operation of a 10~kg detector has been carried out \cite{ds10}, the DarkSide-50 (DS-50) detector has been operating underground in the Hall C of LNGS since 2013.
After a first step with the TPC detector initially filled with the AAr \cite{ds50_aar}, the DS-50 filled with argon derived from underground sources (with a 46~kg fiducial mass TPC) is still currently running after reaching a background-free dark matter search with a sensitivity of 1.14$\times$10$^{-44}$~cm$^{2}$ for a 100~GeV WIMP mass \cite{ds50_uar_532}.\\
The next step of the project foresees to proceed towards a multi-ton detector and a sensitivity improvement of about three orders of magnitude.


\section{The DarkSide-50 detector}
\label{sec:ds50_detector}

\begin{figure}[h!]
\begin{center}
\includegraphics*[width=7.8cm,angle=0]{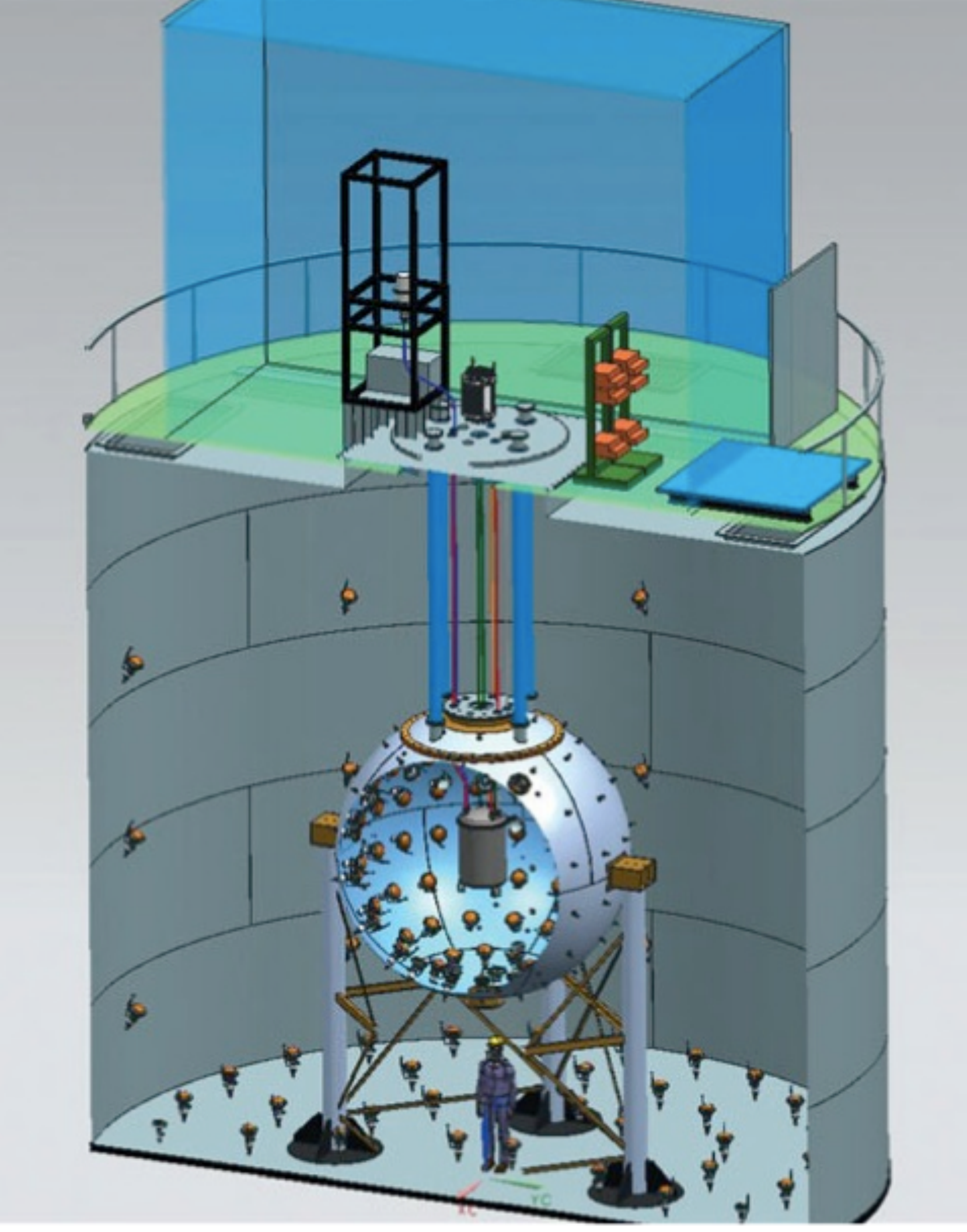}
\includegraphics*[width=6.7cm,angle=0]{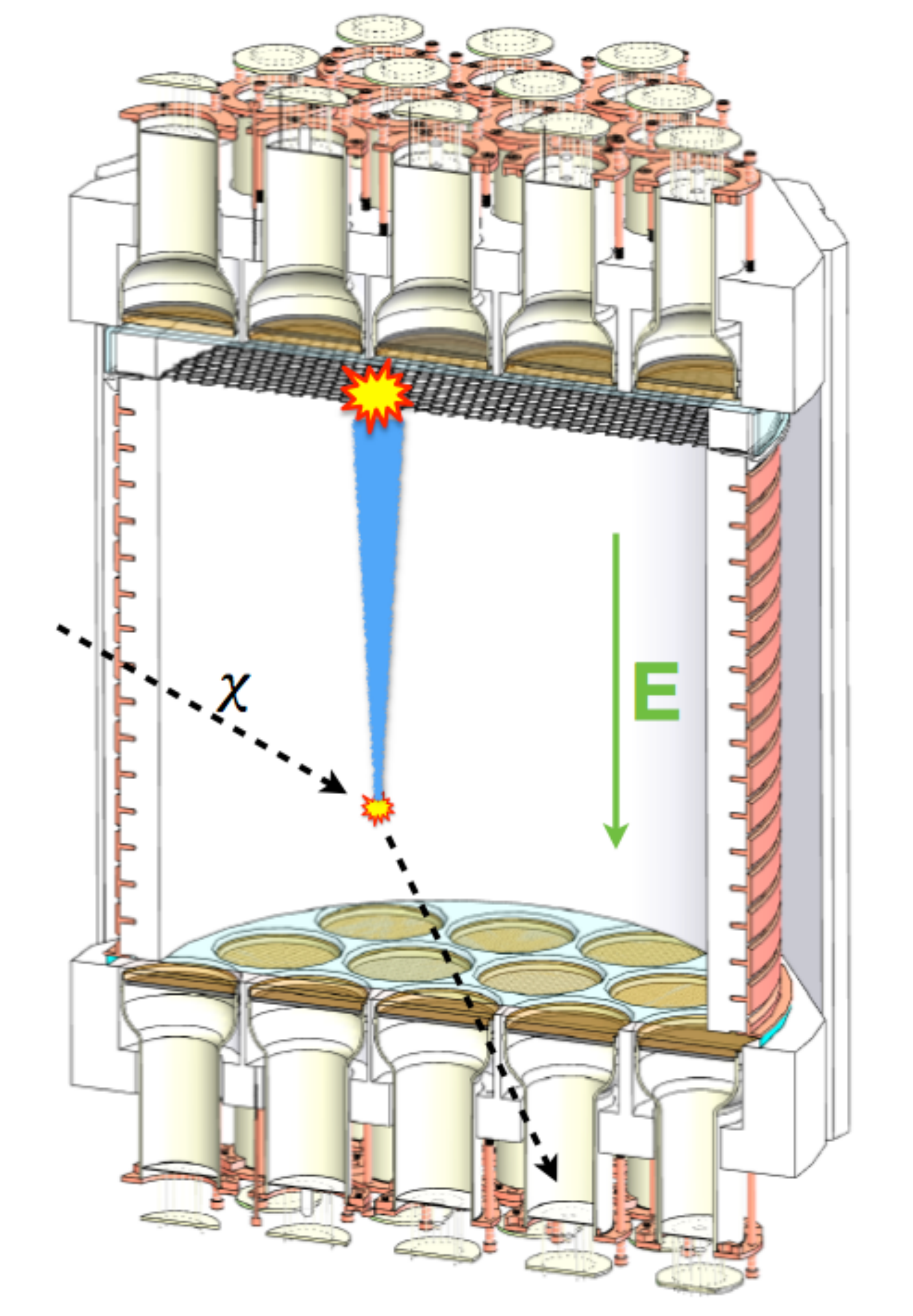}
 \caption{Artistic view of the DS-50 detector with the three nested detectors: TPC in the center of the LSV surrounded by the WCD [Left]; picture of the TPC with the PTFE chamber and the 3'' R11065 PMTs [Right].}
\label{fig:ds50_detector}
\end{center}
\end{figure}

The DarkSide-50 experiment is composed by three nested detectors (see Figure~\ref{fig:ds50_detector} [Left]): innermost is the LAr TPC, acting as dark matter detector containing the liquid argon target, housed by the organic Liquid Scintillator Veto (LSV), serving as shielding and as anti-coincidence for radiogenic and cosmogenic neutrons, $\gamma$-rays and cosmic muons and the most externally Water Cherenkov Detector (WCD), acting as a shield and as anti-coincidence for cosmic muons \cite{ds50_aar, ds50_veto}.\\
The TPC is made by a high reflectivity PTFE square cylinder (36~cm diameter by 36~cm height) with 46~kg active volume, viewed by 38 (19 at bottom and 19 on the top) Hamamatsu HQE R11065 3'' PMTs. The top of the active volume is defined by a stainless steel grid, allowing for setting the drift and extraction fields independently. The gas pocket above the active volume is 1~cm thick. 
Two fused silica windows coated with transparent conductor, ITO, forming the anode and cathode surfaces, are respectively placed at the top and at the bottom of the PTFE chamber. Field shaping copper rings outside the TPC ensure a uniform electric field inside the LAr active volume. 
All inner surfaces are coated with the wavelength shifter tetraphenyl-butadiene (TPB) shifting the VUV 128~nm LAr scintillation light to visible 420~nm light detected by the PMTs (see Figure~\ref{fig:ds50_detector} [Right]).\\
Ionizing particles interacting in the sensitive LAr volume of the detector deposit energy inducing both a prompt scintillation and ionization. The scintillation light represents the primary pulse signal S1. The ionization electrons are drifted upwards by an electric field (200~V/cm) to the liquid surface where they are extracted into a gas pocket by a higher electric field (2.8~kV/cm). The acceleration of the electrons across the gas pocket produces a secondary electroluminescence scintillation signal S2, proportional to the number of ionization electrons. The S1 and S2 pulses enable 3D position reconstruction of the primary interaction site: the time between S1 and S2 gives the vertical position, and the hit pattern of the S2 light on the photosensors gives the transverse position. The 3D position reconstruction allows for rejection of surface backgrounds.\\
The Liquid Scintillator Veto surrounding the TPC is a 4~m diameter stainless steel sphere filled with a mixture of pseudocumene and trimethyl-borate (TMB). The content of $^{10}$B in the scintillator provides a large neutron capture cross section.
The LSV is instrumented with 110 Hamamatsu R5912 8'' PMTs; it acts as an active veto to tag neutrons in the TPC and for allowing for measurement of the neutron background directly in situ.
A Water Cherenkov Detector, made by an 11~m diameter by 10~m height water tank surrounds the LSV; it is equipped with 80 ETL 9351 8'' PMTs and forms an active muon veto to tag cosmogenic induced neutrons. Both the LSV and the WCD provide passive neutron and gamma shielding for the TPC.
%
%
%
%
%
%


\section{DarkSide-50 stability and performances}
\label{sec:calibrations}

The DS-50 detector has been almost continuously running and taking data since 2013.
A good stability of the DS-50 TPC detector has been shown in terms of thermodynamic parameters of cryogenic system, PMTs parameters, electric fields, light yield and liquid argon purity.
An active slow control allows for the set of the cryogenic system serving the TPC detector where the thermodynamic parameters (pressure, argon and nitrogen flows, heating power, argon purification, etc.) are routinely checked, monitored and stored. 
The main feartures of the 38 TPC PMTs are daily checked via laser calibration to estimate the gains (through the single electron response - SER), the SER resolution, the PMTs QE's, rates, voltage supplies and currents.
The evaluation of the TPC light yield (LY) at null field and at 200~V/cm has been performed by using data from a $^{83m}$Kr source with the TPC filled with both atmospheric and underground argon.
In the Figure~\ref{fig:f90} [Left] the resulting spectrum as convolution of $^{39}$Ar and $^{83m}$Kr sources during the AAr campaign has been reported; it has been fit to obtain the measurement of the light yield of the detector at the 41.5~keV reference line of $^{83m}$Kr. The fit of the overall spectrum, accounting for both the $^{39}$Ar and $^{83m}$Kr contributions, returns a light yield, expressed in photoelectron (PE) per unit energy, of (7.9$\pm$0.4)~PE/keV at zero drift field, while a value of (7.0$\pm$0.3)~PE/keV has been measured with the drift field settled at 200~V/cm \cite{ds50_aar}. A slightly improved LY of (8.1$\pm$0.2)~PE/keV was found stable within 2\% during the whole period of operations with the underground argon \cite{ds50_uar_first}.
A very low levels of electronegative impurities in liquid argon have been measured with the $^{83m}$Kr internal calibration, achieving an electron drift lifetime $\tau_{e}$>5~ms during the AAr campaign and reaching an extrapolated value of $\tau_{e}$>11~ms (at 50~V/cm) with the use of UAr.\\
Rates, voltage supplies, currents of the LSV and WCD PMTs are routinely monitored and registered. The LSV PMTs SER have been evaluated by injecting light from a laser.
The light yield of the LSV was measured through the $^{14}$C content in the TMB
and the $^{60}$Co coming from the TPC cryostat stainless steel. By combining the fit results of the two spectra a LSV light yield of (0.54$\pm$0.04)~PE/keV was obtained.\\
A calibration campaign, following the first dark matter search, was started with the aim to perform a characterization of the detector response, a detection efficiency and an energy calibration for both TPC and LSV.
To this purpose, a dedicated calibration hardware was installed in the DS-50 detector, allowing for the placement of several gamma ($^{57}$Co, $^{133}$Ba, $^{137}$Cs, $^{22}$Na) and neutron sources (AmBe, AmC) in the LSV next to the TPC cryostat \cite{calis}.\\
Data acquisition showed an impressive stability reaching a trigger efficiency of $\sim$90\% during the atmospheric argon campaign and up to 95\% with the TPC filled with the UAr. A total amount of more than 1000 live days data have been taken during the AAr and UAr physics runs.\\  
Maintenance of the systems and facilities (cryogenic and gas systems, TPC and vetoes electronics, slow control modules, etc.) related to the experiment have been occasionally performed to keep the detectors in safe conditions and continuously running.


\section{Results with atmospheric argon}
\label{sec:aar_result}

\begin{figure}[h!]
\begin{center}
\includegraphics*[width=7.2 cm,angle=0]{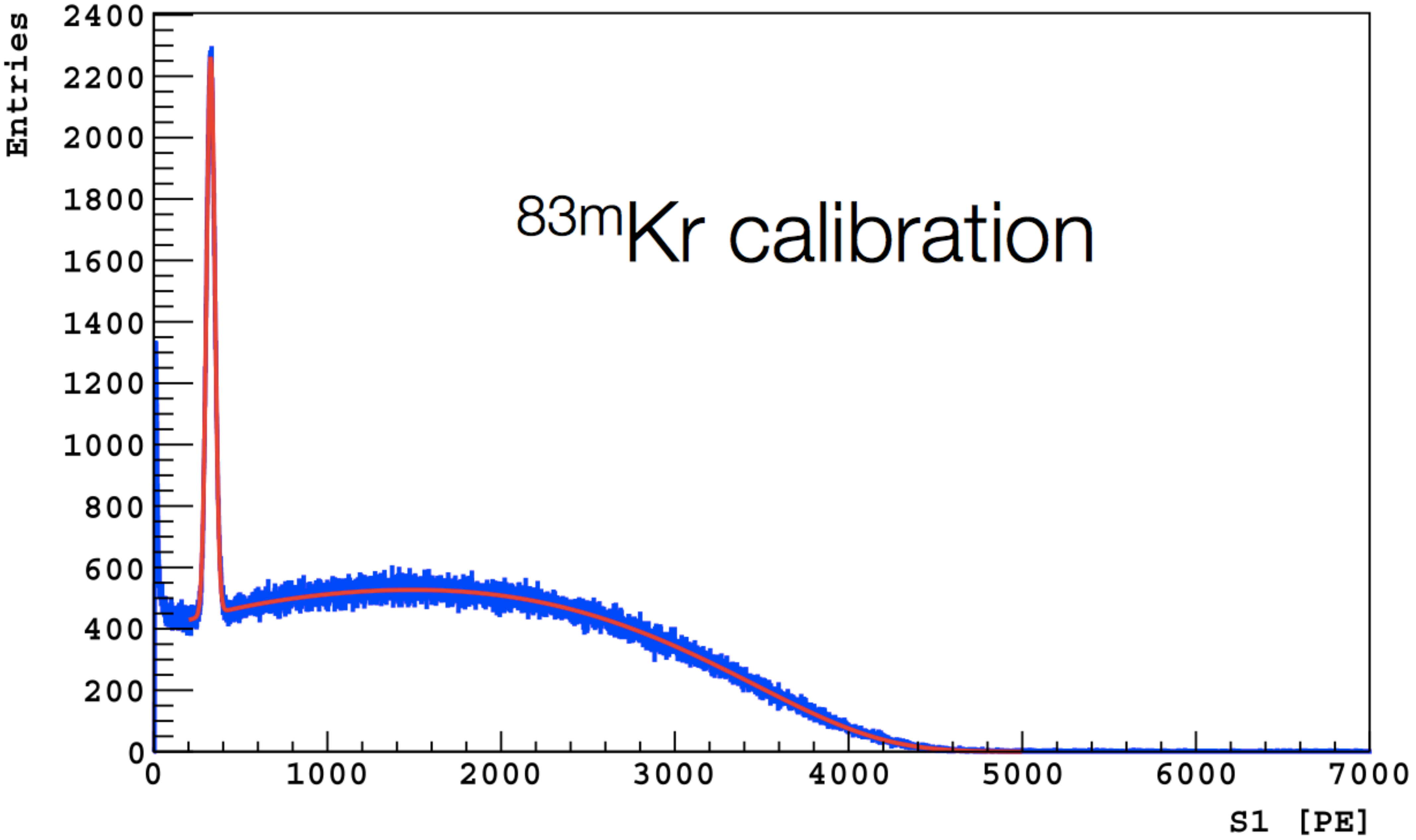}
\includegraphics*[width=7.8 cm,angle=0]{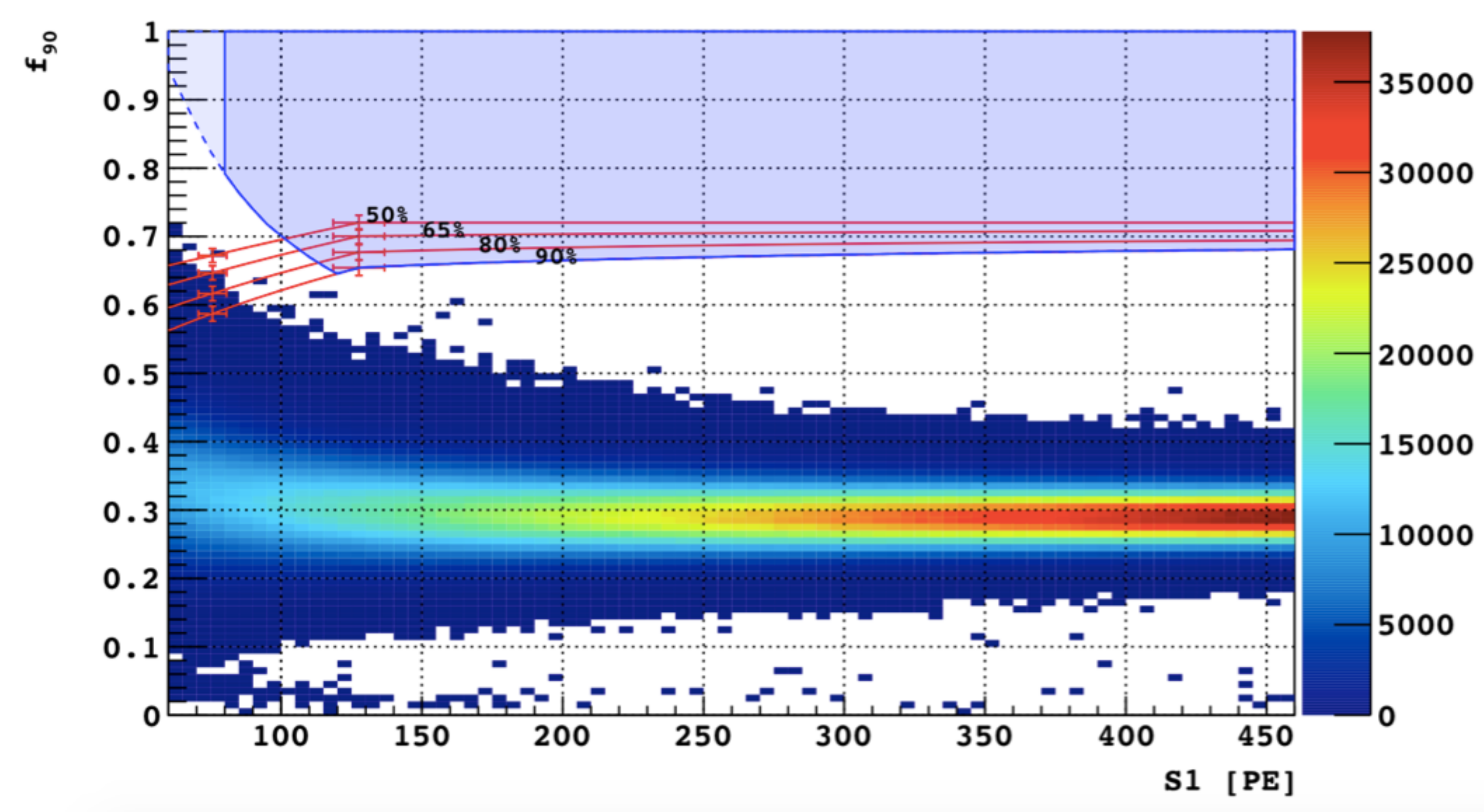}
 \caption{Primary scintillation spectrum from a zero-field run of the DS-50 TPC obtained while the recirculating argon was spiked with $^{83m}$Kr (blue), with the superimposed fit to the $^{83m}$Kr+$^{39}$Ar spectrum (red) [Left]. Distribution of events in the f$_{90}$ vs S1 plane which survive all cuts; shaded blue with solid blue outline: WIMP search region [Right].}
\label{fig:f90}
\end{center}
\end{figure}

A first dark matter search has been performed with data collected with atmospheric argon target with an AAr exposure of 47.1 live days (1422$\pm$67~kg$\times$d exposure) of data acquired in the period October 2013 - May 2014 \cite{ds50_aar}. The TPC trigger rate was dominated by $^{39}$Ar $\beta$-decays, giving a value of 50~Hz.
The $^{39}$Ar background from 47.1 live days of AAr corresponds to that expected in a period $\geq$20 years of UAr DS-50 run at the upper limit $^{39}$Ar activity.
A set of data quality cuts, both for the LAr TPC and the vetoes, were applied. 
A fiducial mass of (36.9$\pm$0.6)~kg was considered. It was defined by requiring the drift time to be between 40~$\mu$s and 334.5~$\mu$s, while no radial cuts were applied.
The WIMP search region was defined in the range of 80$\div$460~PE of S1, corresponding to recoils in the energy range 38$\div$206~keV$_{\bf nr}$. 
No events in the WIMP search region in the f$_{90}$ vs S1 plane resulted in the final dark matter search as reported in Figure~\ref{fig:f90} [Right], leading to the placement of the 90\% CL exclusion curve with a minimum cross section of 6.1$\times$10$^{-44}$~cm$^{2}$ for a 100~GeV/c$^{2}$ WIMP mass. 
%
%
%
%
%
%
%
%
%
%


\section{First result with underground argon}
\label{sec:uar_result}

\begin{figure}[h!]
\begin{center}
\includegraphics*[width=7.5 cm,angle=0]{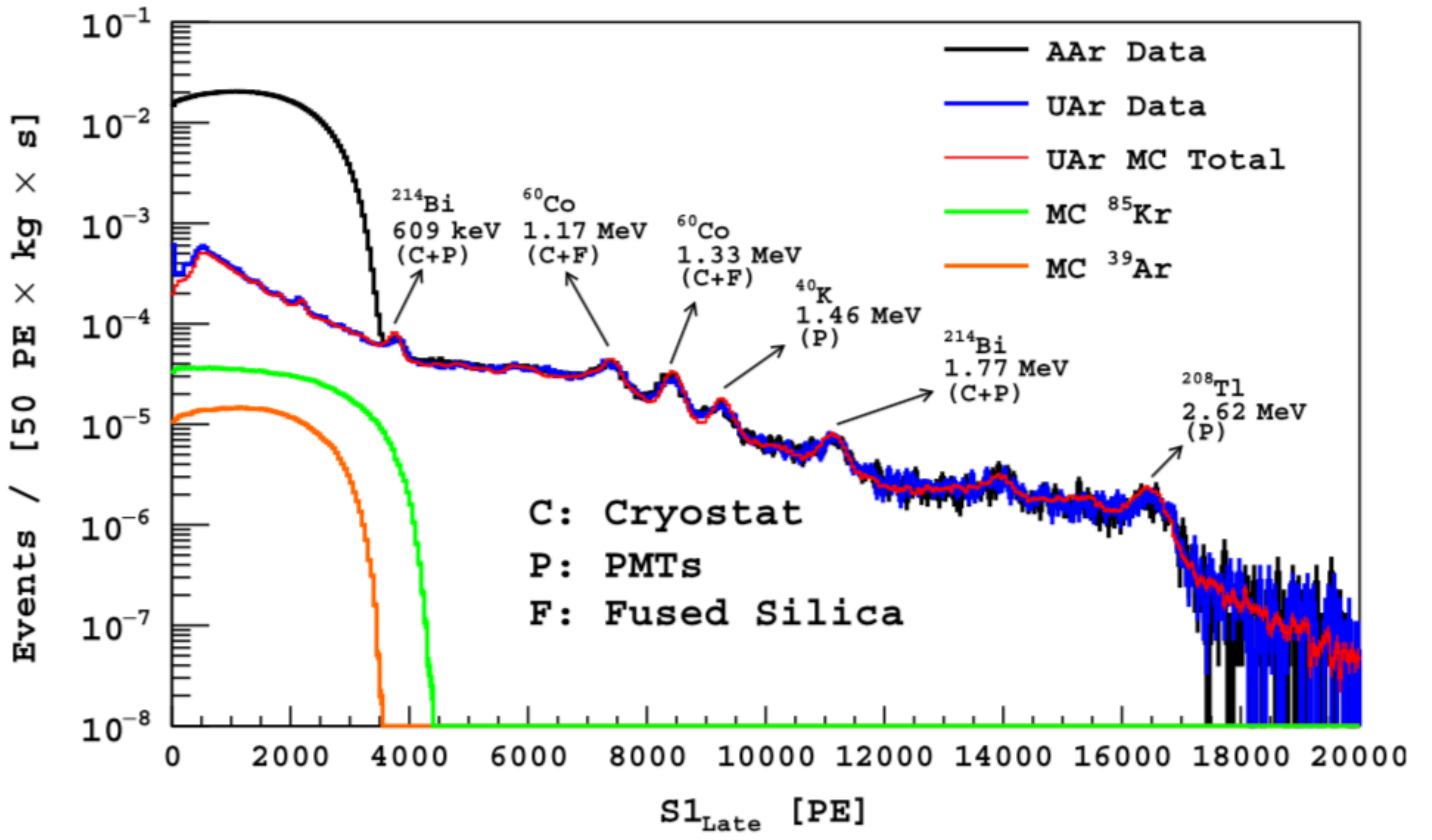}
\includegraphics*[width=7.5 cm,angle=0]{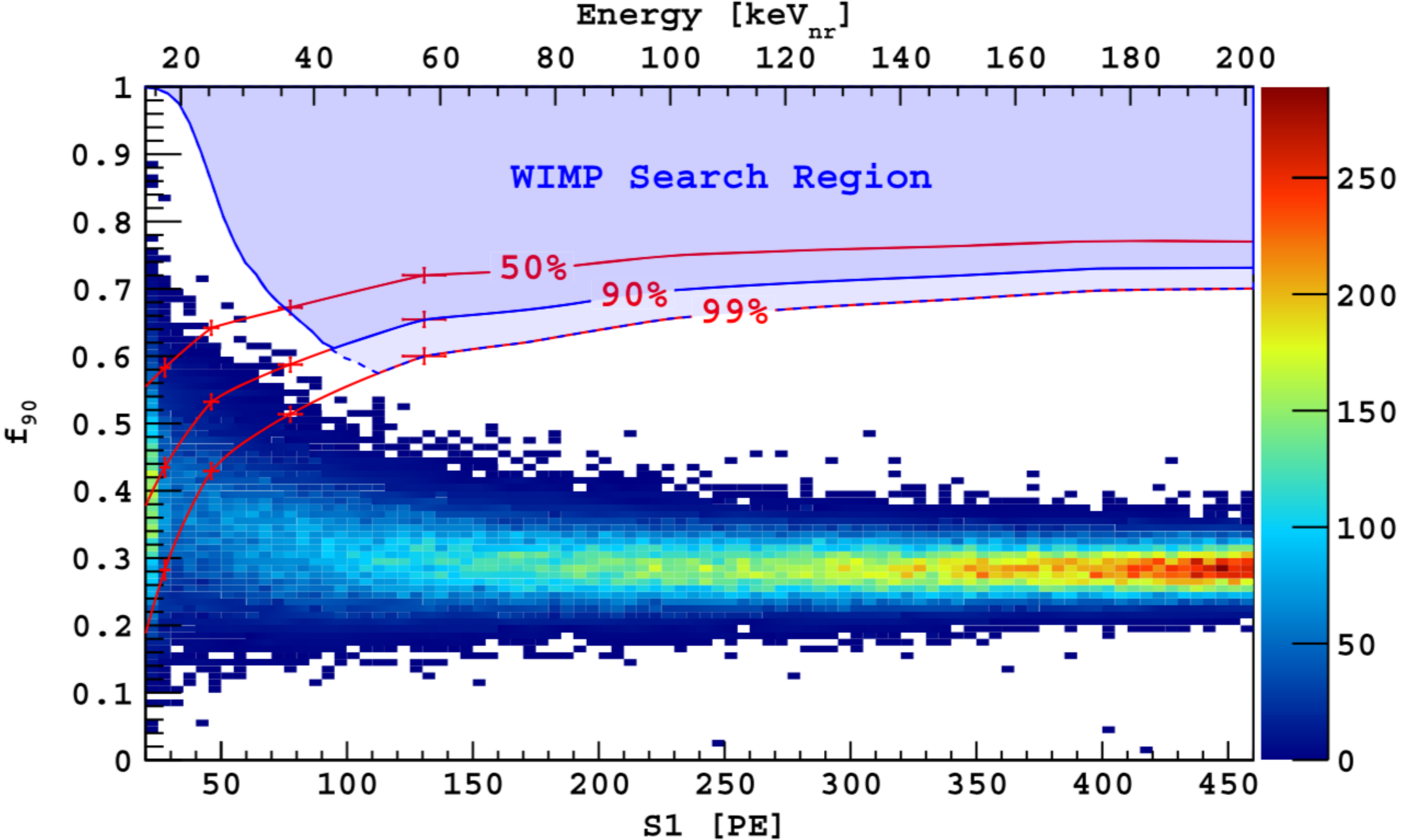}
\caption{Comparison of the measured field off spectra for the UAr (blue) and AAr (black) targets, normalized to exposure [Left]. 
Distribution of events in the f$_{90}$ vs S1 plane surviving all cuts in the energy region of interest; shaded blue with solid blue outline: WIMP search region [Right].}

\label{fig:uar_spectrum}
\end{center}
\end{figure}

After demonstrating the first results with the atmospheric argon, the DS-50 TPC was emptied of 
AAr in March 2015 and refilled with the UAr; data taking started on April 1, 2015 and over 70.9 live days (2616$\pm$43~kg$\times$d) have been accumulated \cite{ds50_uar_first}.
Data acquired with the UAr showed a trigger rate of $\sim$1.5~Hz, significantly reduced if compared with the AAr trigger rate of $\sim$50~Hz. 
After a minimal set of cuts for single scatter events selection in the TPC was applied a good agreement between the rate of events beyond the $^{39}$Ar endpoint for the AAr and UAr spectra was observed, as reported in Figure~\ref{fig:uar_spectrum} [Left], indicating an unchanged light yield after TPC chamber was filled with the UAr. Further calibration with the $^{83m}$Kr source confirmed an improved light yield compared to the AAr.
A MC simulation of the DarkSide-50 detectors also developed to compare the ER background from UAr with that from AAr was able to determine the $^{39}$Ar and $^{85}$Kr activities in the UAr to be (0.73$\pm$0.11)~mBq/kg and (2.05$\pm$0.13)~mBq/kg, respectively. 
These results lead to a $^{39}$Ar activity of the UAr corresponds to a reduction by a factor of (1.4$\pm$0.2)$\times$10$^{3}$ compared to AAr.
Then a WIMP search region for the UAr has been defined in the f$_{90}$ vs S1 plane having an energy region of interest of 20$\div$460~PE in S1 (13$\div$201~keV$_{\bf nr}$). No events resulted in the WIMP search region.
The null result of the UAr exposure gave an upper limit on the WIMP-nucleon spin-independent cross section of 3.1$\times$10$^{-44}$~cm$^{2}$  for a WIMP mass of 100~GeV/c$^{2}$ and un upper limit of 2.0$\times$10$^{-44}$~cm$^{2}$  for a WIMP mass of 100~GeV/c$^{2}$ has been obtained combining the UAr results with the one of the AAr exposure.
%
%
%
%
%
%
%


\section{DarkSide-50 physics results}
\label{sec:ds-50_results}

\begin{figure}[h!]
\begin{center}
\includegraphics*[width=7.5 cm,angle=0]{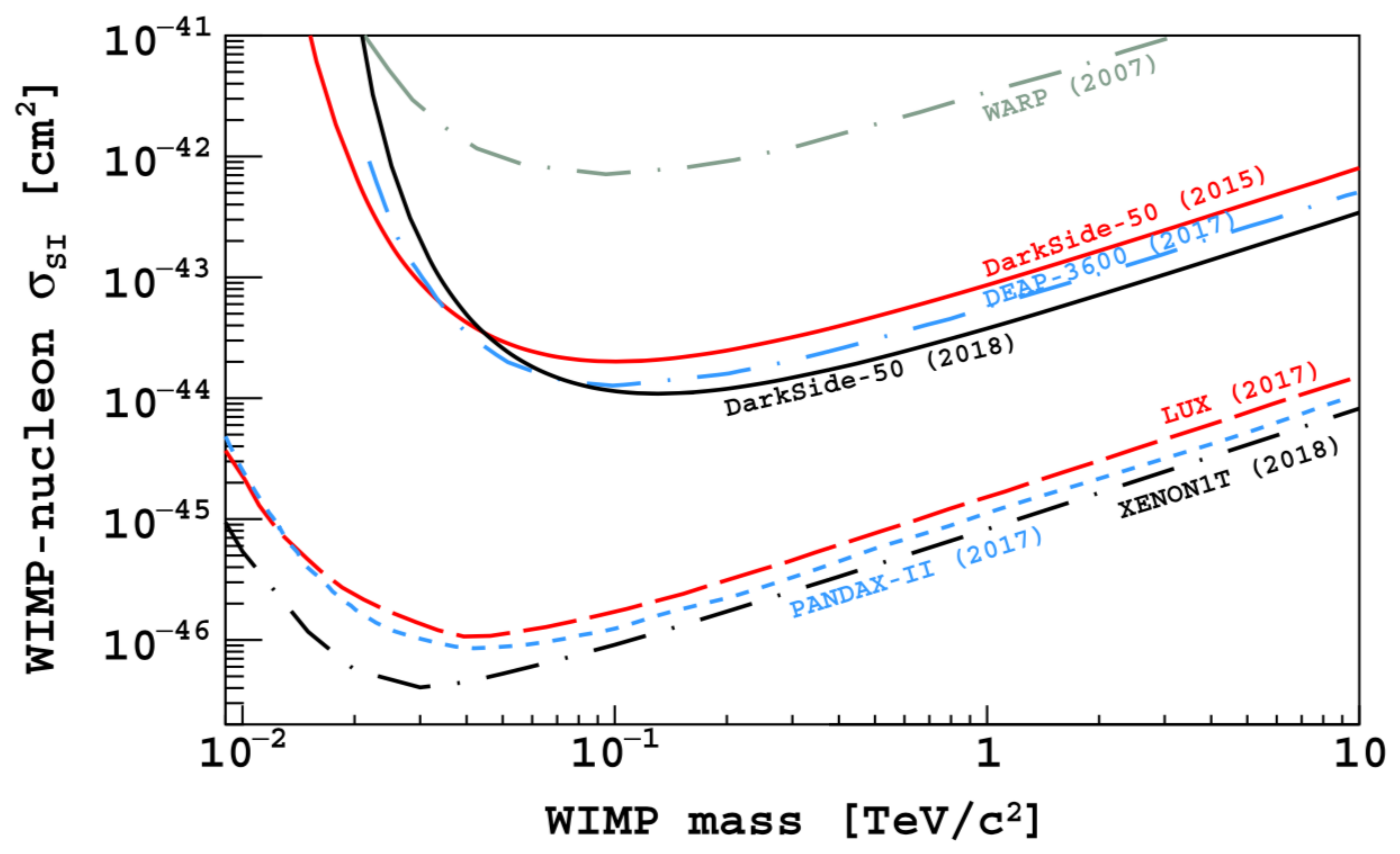}
\includegraphics*[width=7.5cm,angle=0]{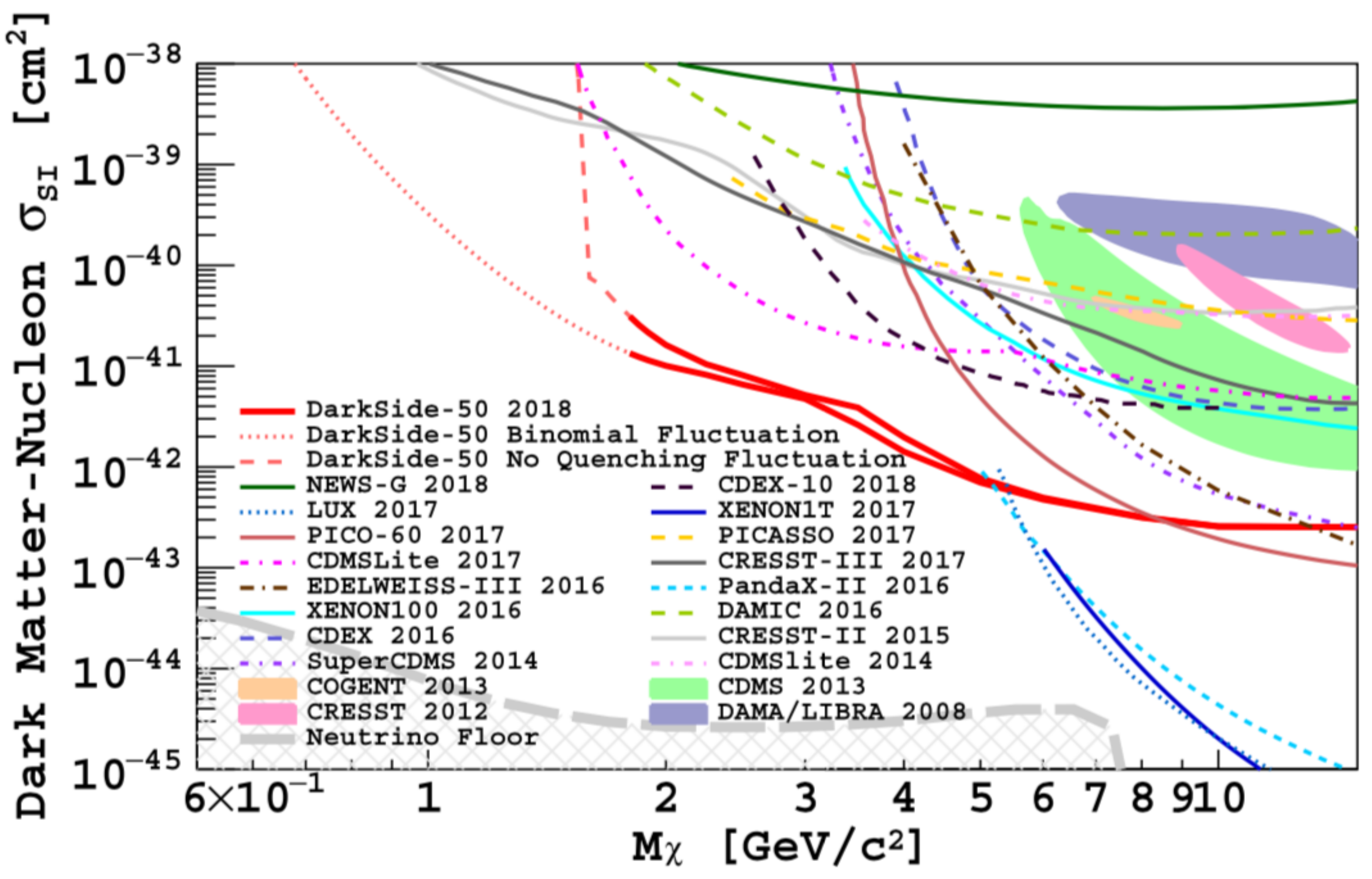}
 \caption{Spin-independent DM-nucleon cross section 90\% C.L. exclusion limits [Left]. Upper limits (90\%) on spin-independent DM-nucleon cross sections from DS-50 in the range above 1.8~GeV/c$^{2}$ [Right].}
\label{fig:uar_physics}
\end{center}
\end{figure}

After a first UAr campaign has been performed a new set of data consisting of 532.4 live days (16660$\pm$270~kg$\times$d) with the underground argon (from August 2, 2015, to October 4, 2017) has been taken \cite{ds50_uar_532}.
A blind analysis with the aim to design a set of criteria for the background rejection without prior inspection of events in the final search region has been carried out. An expected background of 0.1 event as an acceptable level was chosen.
Data were initially checked run-by-run removing those runs affected by hardware and software issues.
Basic cuts for the event quality selection and against cosmic ray activation have been applied, backgrounds from surface events, radiogenic and cosmogenic neutrons have been rejected, electron recoil backgrounds mainly due to scintillation and Cherenkov events in the TPC materials have been evaluated, fiducial cut accounted for a fiducial mass of 31.3$\pm$0.5~kg, then the background surviving to the cuts was estimated. The blind data selection was followed by an unblind analysis.
No events were observed in the predefined DM search region and as reported in Figure~\ref{fig:uar_physics} [Left] an upper limit on the spin-independent DM-nucleon cross section at 1.14$\times$10$^{44}$~cm$^{2}$ (3.78$\times$10$^{44}$~cm$^{2}$, 3.43$\times$10$^{43}$~cm$^{2}$) for 100~GeV/c$^{2}$ (1~TeV/c$^{2}$, 10~TeV/c$^{2}$) DM particles was found.\\
A subset of of 532.4 live days data for a global amount of 6786.0~kg$\times$d exposure was used for an analysis based on the ionization signal with the purpose to explore the dark matter mass range below 20~GeV/c$^{2}$ \cite{ds50_low_mass, ds50_low_mass_constraints}.
Some quality cuts were applied to the data before analysis. A dedicated MC simulation reproducing the spatial distribution of S2 light was used to estimate the acceptance of all the applied cuts and rejected events.
The Ds-50 TPC was found fully efficient at 0.1~keV$_{\bf ee}$ for the ionization only signals. 
A 90\% C.L. exclusion limit above 1.8~GeV/c$^{2}$ for the spin-independent cross section of dark matter WIMPs on nucleons, extending the exclusion region for dark matter below previous limits in the range 1.8$\div$6~GeV/c$^{2}$ was obtained. Results achieved with the ionization S2 signal in DS-50 are reported in Figure~\ref{fig:uar_physics} [Right].
%
%
%
%
%
%
%
%
%


\section{Conclusions}
\label{sec:concl}

The DarkSide-50 experiment is stably and continuously operating underground at LNGS since October 2013. 
The first dark matter search campaign of DS-50, with the TPC filled with the atmospheric argon, set the most stringent limit on the WIMP-nucleon cross section for a liquid argon target.
The DarkSide-50 detector, with the TPC filled with UAr, is currently operating and accumulating exposure in a stable, low-background configuration. 
A depletion factor of $^{39}$Ar activity in the UAr resulted in a reduction of the $^{39}$Ar concentration of $\sim$1400.\\
Results with an exposure of 532 days of UAr have been obtained leading to a sensitivity of 1.14$\times$10$^{-44}$~cm$^{2}$  for a WIMP mass of 100~GeV/c$^{2}$.
At the same time a very promising result based on the ionization-only signal has been
achieved for the low mass dark matter search exploring the region above 1.8~GeV/c$^{2}$.\\
DS-50 experiment with UAr technology demonstrated the feasibility of large scale LAr detectors for direct dark matter search and based on the DS-50 experience LAr detectors for direct DM search will be scaled to multi-hundred ton representing the most powerful background-free technique for the ultimate experiment in dark matter search.
%



\acknowledgments

The  DarkSide  Collaboration  offers  its  profound gratitude to the LNGS and its staff  or their  invaluable  technical and logistical support. We also thank the Fermilab Particle Physics, Scientific, and Core Computing Divisions.  
Construction and operation of the DarkSide-50 detector was supported by the U.S. National Science Foundation (NSF) (Grants PHY-0919363, PHY-1004072, PHY-1004054,
PHY-1242585, PHY-1314483, PHY-1314501, PHY-1314507, PHY-1352795, PHY-1622415, and  associated collaborative grants PHY-1211308 and PHY-1455351), the Italian Istituto Nazionale di Fisica Nucleare, the U.S. Department  of  Energy  (Contracts DE-FG02-91ER40671,  
DE-AC02-07CH11359, and DE-AC05-76RL01830), the Russian Science Foundation (Grant 16-12-10369), the Polish NCN (GrantUMO-2014/15/B/ST2/02561) and the Foundation for Polish Science (Grant Team2016-2/17).  
We also acknowledge financial support from the French Institut National de Physique Nucl\'eaire et de Physique des Particules (IN2P3), from the UnivEarthS Labex program of Sorbonne Paris 
Cit\'e  (Grants  ANR-10-LABX-0023 and ANR-11-IDEX-0005-02), and from the S\~ao Paulo Research Foundation (FAPESP) (Grant 2016/09084-0).



\end{document}